\documentclass[aps,prd,twocolumn,showpacs,groupedaddress]{revtex4}
\usepackage{epsfig}
\usepackage{longtable}
\usepackage{amsfonts}

\usepackage{amsmath}
\usepackage{amssymb}

\newcommand{\be}{\begin{equation}}
\newcommand{\ee}{\end{equation}}
\begin{document}

\title{Limits to differences in active and passive charges
}

\author{C.\ L{\"a}mmerzahl$^1$, A. Macias$^2$, and H. M\"uller$^3$}
\address{$^{1}$ZARM, University of Bremen, Am Fallturm, 28359 Bremen, Germany \\
$^{2}$Dept. de F\'ysica, Universidad Aut\'onoma Metropolitana Iztapalapa, A.P. 55–534, C.P. 09340, M\'exico, D.F., M\'exico \\
$^3$Physics Department, Stanford University, Stanford, CA 94305-4060, USA}
\date\today
\begin{abstract}
We explore consequences of a hypothetical difference between
active charges, which generate electric fields, and passive
charges, which respond to them. A confrontation to experiments
using atoms, molecules, or macroscopic matter yields limits on
their fractional difference at levels down to $10^{-21}$, which at
the same time corresponds to an experimental confirmation of
Newtons third law.
\end{abstract}


\maketitle

\section{Introduction}

In electrodynamics, one may distinguish between two types of
charge: The {\em active charge} $q_{\rm a}$ is the source of the
electric field,
\begin{equation}
\mbox{\boldmath$\nabla$} \cdot \mbox{\boldmath$E$} = 4 \pi q_{\rm a} \delta(x) \,,
\end{equation}
whereas the {\em passive charge} $q_p$ reacts to it:
\begin{equation}
m \ddot{\mbox{\boldmath$x$}} = q_{\rm p} \mbox{\boldmath$E$} \, .
\end{equation}
Here, $m$ is the inertial mass and ${\mbox{\boldmath$x$}}$ the
position of the particle. In gravitational physics, a hypothetical
difference between active and passive gravitational mass has been
considered and confronted with laboratory and astrophysical
observations. However, as yet nothing similar has been done for
the electric and magnetic analogue, in spite of a long history of
precision experiments that includes tests of the $1/r$ Coulomb
potential, searches for an electrostatic fifth force
\cite{FischbachTalmadge99,SpavieriGilliesRodriguez04}, the photon
mass \cite{TuLuoGillies05}, and violations of Lorentz invariance
\cite{Amelino-Cameliaetal05,Mattingly05}. Current Maxwell theory
tacitly assumes the equality of passive and active charges and is
fundamental to a broad range of theoretical and experimental
physics; any inequality would thus have serious consequences
throughout science, from the  standard model of particle physics
to practical applications like precision metrology. --- Here, we
show that limits as low as $10^{-21}e$ can be derived for protons
and electrons by introducing the concepts of active and passive
neutrality. Furthermore, we identify signatures for such a
difference in atomic spectroscopy. We extend the analysis to
active and passive magnetic moments and find corresponding limits
from hyperfine spectroscopy. Our limits appear important in the
context of recent quantum gravity scenarios, where all sorts of
symmetries (like Lorentz and CPT invariance or the equivalence
principle) are expected to be violated
\cite{Amelino-Cameliaetal05}.

\subsection{Model}

The dynamics of two particles located at
${\mbox{\boldmath$x$}}_{1,2}$ in their mutual electric fields are
described by the equations
\begin{eqnarray}\label{eqofmotion}
m_1 \ddot{\mbox{\boldmath$x$}}_1 &=& q_{\rm 1p} q_{\rm 2a}
\frac{{\mbox{\boldmath$x$}}_2 -
{\mbox{\boldmath$x$}}_1}{|{\mbox{\boldmath$x$}}_2 -
{\mbox{\boldmath$x$}}_1|^3} + q_{\rm 1p}
\mbox{\boldmath$E$}({\mbox{\boldmath$x$}}_1) \, ,\nonumber \\
m_2 \ddot{\mbox{\boldmath$x$}}_2 &=& q_{\rm 2p} q_{\rm 1a}
\frac{{\mbox{\boldmath$x$}}_1 -
{\mbox{\boldmath$x$}}_2}{|{\mbox{\boldmath$x$}}_1 -
{\mbox{\boldmath$x$}}_2|^3} + q_{\rm 2p}
\mbox{\boldmath$E$}({\mbox{\boldmath$x$}}_2) \, ,
\end{eqnarray}
where $\mbox{\boldmath$E$}$ denotes a homogenous external electric
field, and $q_{\rm 1p}, q_{\rm 1a}, q_{\rm 2p}, q_{\rm 2a}$ are
the respective passive and active charges. (This non-relativistic
description will be sufficient for our purpose, which is to
identify stringent constraints on any inequality of active and
passive charges from experiments. In the light of these limits, it
is not necessary to consider the relativistic equations of
motion.) For the equation of motion of the center of mass
$\mbox{\boldmath$X$}$, we find
\begin{equation}
\ddot{\mbox{\boldmath$X$}} = \frac{q_{\rm 1p} q_{\rm 2p}}{M}
C_{21} \frac{\mbox{\boldmath$x$}}{|\mbox{\boldmath$x$}|^3} +
\frac{1}{M} \left(q_{\rm 1p} + q_{\rm 2p}\right)
\mbox{\boldmath$E$} \label{centerofmassmotion}
\end{equation}
where $M = m_1+m_2$, $\mbox{\boldmath$x$}$ is the relative
coordinate, and
\begin{equation}
C_{21} = \frac{q_{\rm 2a}}{q_{\rm 2p}} - \frac{q_{\rm 1a}}{q_{\rm 1p}} \, .
\end{equation}
Thus, if active and passive charges are different, the center of
mass shows a self--acceleration along the direction of
$\mbox{\boldmath$x$}$, in addition to the acceleration caused by
the external field ${\mbox{\boldmath$E$}}$. This can be
interpreted as a violation of Newton's third law {\em actio}
equals {\em reactio} for electric forces. $C_{21} = 0$ means that
the ratio between the active and passive charge is the same for
both particles. If this ratio is the same for all particles, then
it can be absorbed into a redefinition of the electric charges and
has no observable consequences. The dynamics of the relative
coordinate is given by
\begin{eqnarray}
\ddot{\mbox{\boldmath$x$}} 
& = & - \frac{1}{m_{\rm red}} q_{\rm 1p} q_{\rm 2p} D_{21}
\frac{\mbox{\boldmath$x$}}{|\mbox{\boldmath$x$}|^3} \, , \label{RelativeMotion}
\end{eqnarray}
where
\begin{equation}
D_{21}=\frac{m_1}{M} \frac{q_{\rm 1a}}{q_{\rm 1p}} + \frac{m_2}{M}
\frac{q_{\rm 2a}}{q_{\rm 2p}}= \frac{q_{\rm 1a}}{q_{\rm 1p}} +
\frac{m_2}{M} C_{21}
\end{equation}
and $m_{\rm red}$ is the reduced mass. In the standard framework,
$D_{21}=1$. Bound solutions of the equation of motion
(\ref{RelativeMotion}) are ellipses; choosing suitable
coordinates, they can be normal to the $z-$axis. The simplest case
is circular motion, where $\mbox{\boldmath$x$}(t) = x_0(
\cos(\omega t),\sin(\omega t),0)$. The center of mass oscillates
at a frequency $\omega$, which is related to the energy of the
system. The acceleration of the center of mass vanishes on
average, $\langle \ddot{\mbox{\boldmath$X$}} \rangle = 0$. Thus,
it is not necessarily observable. These considerations extend to
many particle systems, e.g., to atoms having many electrons.

Of course, a large $C_{21}$ would likely be observed routinely in
chemistry and physics. For example, the Born-Mayer model predicts
the bond energy of ionic crystals to 1-10\% accuracy, so
$C_{21}\gtrsim 1\%$ could be noticed; also a change of bond
lengths would result \cite{solidstate} and be detectable at this
level. The limit to the accuracy is by the complex nature of the
crystals. However, the experiments to be discussed below can
provide sensitivity up to 19 orders of magnitude better.

\subsection{Active and passive mass}

The analogous case of active and passive {\it masses} has first
been discussed by Bondi \cite{Bondi57}: The equations of motion
for a gravitationally bound two body system have the same
structure as for electric bound charges:
\begin{eqnarray}
\ddot{\mbox{\boldmath$x$}}_1 = G\frac{m_{\rm 1p} m_{\rm
2a}}{m_{\rm 1} } \frac{{\mbox{\boldmath$x$}}_2 -
{\mbox{\boldmath$x$}}_1}{|{\mbox{\boldmath$x$}}_2 -
{\mbox{\boldmath$x$}}_1|^3} \,,\nonumber \\
 \ddot{\mbox{\boldmath$x$}}_2 = G\frac{m_{\rm 2p} m_{\rm 1a}}{m_{\rm 2} }
 \frac{{\mbox{\boldmath$x$}}_1 - {\mbox{\boldmath$x$}}_2}{|{\mbox{\boldmath$x$}}_1 -
 {\mbox{\boldmath$x$}}_2|^3}\,,
\end{eqnarray}
where the indices p and a denote the passive and active
gravitational mass, respectively, and $G$ is the gravitational
constant. (As throughout, $m$ without these indices denotes the
inertial mass.) Thus, an inequality of active and passive masses
results in a self--acceleration of the center of mass if $\bar
C_{21} = (m_{\rm 2a}/m_{\rm 2p})-(m_{\rm 1a}/m_{\rm 1p})\neq 0$,
which again can be interpreted as violation of Newton's third law
for gravitational forces. A limit has been derived by lunar laser
ranging: that no self--acceleration of the moon has been observed,
yields a limit of $|\bar C_{\rm Al-Fe}| \leq 7 \cdot 10^{-13}$
\cite{BartlettvanBuren86}. The dynamics of the relative coordinate
\begin{equation}
\ddot{\mbox{\boldmath$x$}} = - G \frac{m_{\rm 1p} m_{\rm 2p}}{m_1
m_2} \left(m_1 \frac{m_{\rm 1a}}{m_{\rm 1p}} + m_2 \frac{m_{\rm
2a}}{m_{\rm 2p}}\right)
\frac{\mbox{\boldmath$x$}}{|\mbox{\boldmath$x$}|^3} \, .
\end{equation}
has been probed in a laboratory experiment by Kreuzer
\cite{Kreuzer68} with the result $|\bar C_{21}| \leq 5 \cdot
10^{-5}$. Note that these experiments are purely gravitational
ones. For the astrophysical observations, this is because
astronomical bodies do not carry active electric charges.
(Otherwise, they would attract passively charged particles. If
these carry active charges of the same sign, this will eventually
neutralize the active charge.) In laboratory experiments, electric
neutrality is ensured by grounding. These experiments thus confirm
the equality of active and passive mass, independent of any
inequality between active and passive electric charge.

Owing to the extreme relative weakness of the gravitational force
compared to the electrical one, it would take a huge violation of
the equality of active and passive mass to mimic a signal for an
inequality of active and passive charge. For example, the
gravitational force between the electron and the proton in a
hydrogen atom is $\sim 10^{-39}$ times the electrical one. Thus,
to mimic an electrical $C_{21}\sim 10^{-20}$ in the hydrogen
spectroscopy experiments discussed later, it would take a
gravitational $\bar C_{21} \sim 10^{+19}$. A $\bar C_{21}$ of this
magnitude is clearly ruled out. Other experiments discussed by us
are based on measuring the active charge of a macroscopic number
of atoms by use of an electrometer. The electrometer is based on
the Coulomb force caused by the charge to be measured on a number
of electrons. Again, the gravitational interaction with these
electrons is much weaker than the electrical and it would take a
huge gravitational $\bar C_{21}$ to mimic an electrical $C_{21}$.
Because of this we can neglect any inequality of gravitational
mass (as well as standard gravitational effects) in the remainder
of this paper. The limits we will find are thus independent of an
inequality of active and passive mass.

Moreover, since the acceleration of charges is proportional to
$q_{\rm p}/m$ (where $m$ is the inertial mass), one might ask
whether measurements may be insensitive to changes in the passive
charge that are accompanied by proportional changes in the
inertial mass. This question is best answered by an explicit
example. We shall do so in section \ref{spectroscopy}.

\subsection{Comparison of the gravitational and electrical case}

The electric case differs from the gravitational one in three
important ways: (i) In the gravitational case, the weak
Equivalence Principle $m = m_{\rm p}$ implies that paths of
particles depend on the active gravitational mass only. (ii) Since
the timescale of electric phenomena is much shorter than that of
gravitational ones, the motion of the center of mass cannot be
monitored. This kind of test is therefore not at our disposal.
(iii) Contrary to the gravitational case, electric charges can
have different signs. Therefore, we can define {\em active}
neutrality $q_{\rm 1a} + q_{\rm 2a} = 0$ as well as {\em passive}
neutrality $q_{\rm 1p} + q_{\rm 2p} = 0$. This allows us to find
alternative tests of the equality of active and passive charges:
An actively neutral system may not be passively neutral and vice
versa. Both definitions of neutrality are compatible, but a system
can be actively and passively neutral if and only if $C_{21} = 0$.
Therefore, a self--acceleration of the center of mass occurs only
if the system possesses a nonzero total active or passive charge.

\section{Experiments with macroscopic matter}

In order to interpret tests of the neutrality of atoms and
molecules as tests of the equality of active and passive charge,
we study compound particles that are actively or passively
neutral. If we assume passive neutrality, there will be no
acceleration due to the external field ${\mbox{\boldmath$E$}}$ and
$C_{21}$ reduces to $(q_{\rm 2a} + q_{\rm 1a})/q_{\rm 2p}$. The
difference between active and passive charges is now related to
the active neutrality of the composed system: a passively neutral
system may still generate an electric field according to
\begin{equation}
\phi(\mbox{\boldmath$x$}) = \frac{q_{\rm 1a}}{|\mbox{\boldmath$x$} -
{\mbox{\boldmath$x$}}_1|} + \frac{q_{\rm 2a}}{|\mbox{\boldmath$x$} -
{\mbox{\boldmath$x$}}_2|} = \frac{q_{\rm 1a} + q_{\rm 2a}}{|\mbox{\boldmath$x$}|} +
\ldots \approx C_{21} \frac{q_{\rm 2p}}{|\mbox{\boldmath$x$}|}
\end{equation}
(where the dots denote dipole and higher order multipole
contributions that are neglected here). On the other hand, an
actively neutral system in an homogenous external electric field
feels a force
\begin{equation}
M \ddot{\mbox{\boldmath$X$}} = (q_{\rm 1p} + q_{\rm 2p})
\mbox{\boldmath$E$} = \frac{q_{\rm 2p}}{q_{\rm 2a}} q_{\rm 1a}
C_{12} \mbox{\boldmath$E$}\,.
\end{equation}
Thus, we can distinguish two types of tests of neutrality: (i)
Tests of active neutrality, which measure the electric monopole
field created by a passively neutral system, and (ii) tests of
passive neutrality, which measure the force imposed by an external
field onto an actively neutral system.

A review of tests of the neutrality of atoms can be found in
\cite{UnnikrishnanGillies04}. One type of experiments, gas efflux
experiments, tests the active neutrality. They are based on
observing the charge of a metallic container during an in- or
outflow of gas or liquid. [The charge measurement is based on the
electric field caused by the charge and is therefore sensitive to
active charge only. This also applies to the modern electrometers
that use field-effect transistors.] With each of $N$ atoms or
molecules containing a number $n_p$ of protons and electrons and
$n_n$ neutrons, the charge $N[n_p(q_{e,\rm a}+q_{p,\rm a})+n_n
q_{n,\rm a}]$ is measured. The indices $e,p,$ and $n$ denote the
electron, the proton, and the neutron. An interesting modern
variant \cite{Classenetal98} uses superfluid He as a medium.

The passive neutrality has been tested by a variety of methods,
see Ref. \cite{UnnikrishnanGillies04} for details: (i) Levitation
experiments \cite{MarinelliMorpurgo82} follow the famous
experiment by Millikan for the measurement of the electric charge
of atoms. (ii) In acoustic resonator experiments
\cite{DyllaKing73}, one applies an alternating electric field
within an acoustic resonator and listens for the sound that would
result due to a passively charged medium. (iii) Atom
\cite{Hughes57} or neutron \cite{Baumannetal88} beam experiments
measure the deflection of a beam of atoms or neutrons which
transverses an electric field.

The limits assembled in Tab. \ref{atomtests} are at levels down to
$10^{-21}$ elementary charges for the active and passive charge of
various combinations of electrons, protons, and neutrons. If we
assume that there are no cancellations, we can thus conclude that
$|C_{pe}| \leq 10^{-21}$, which can be regarded as a verification
of the Newton's third law for electric forces at the $10^{-21}$
level.

\begin{table}
\centering \caption{\label{atomtests} Various tests of the
neutrality of atoms. If no particle is specified, $q_{\rm p}$
refers to the passive charge of the atoms or molecules used in the
experiment, divided by the charge number of that particle (and
analogous for $q_{\rm a}$).}
\begin{tabular}{llc}\hline
Method & Limit $/(10^{-20}e)$\\ \hline
Gas efflux (350\,g CO$_2$) \cite{PiccardKessler25} & $q_{p, {\rm a}}-q_{e, {\rm a}}=0.1(5)$ \\
Gas efflux (Ar/N) \cite{HillasCranshaw60} & $q_{{\rm H}, {\rm a}}=1(3)$; $q_{n, {\rm a}}=-1(3)$ \\
Gas efflux \cite{King60} & $q_{{\rm He, a}}=-4(2)$ \\
Superfluid He \cite{Classenetal98} & $q_{{\rm He, a}}=-0.22(15)$ \\
Levitator \cite{MarinelliMorpurgo82} & $|q_{\rm p}|\lesssim 1000$ \\
Acoustic resonator (SF$_6$) \cite{DyllaKing73} & $|q_{\rm p}|\leq 0.13$ \\
Cs beam \cite{Hughes57} & $q_{\rm p}=90(20)$ \\
Neutron beam \cite{Baumannetal88} & $q_{n, {\rm p}}=-0.4(1.1)$ \\ \hline
\end{tabular}
\end{table}

\section{Spectroscopy}
\label{spectroscopy}

We now study the shift of atomic transition frequencies due to an
inequality of active and passive charges. Although the center of
mass motion of the two--particle system cannot be quantized in
general, the relative motion can. The Hamiltonian is
\begin{equation}
H = \frac{{\mbox{\boldmath$p$}}^2}{2 m_{\rm red}} + D_{21} \frac{q_{\rm 1p} q_{\rm 2p}}{|\mbox{\boldmath$x$}|} \, .
\end{equation}
The energy levels for a single electron atom are proportional to
the square of a modified fine structure constant
\begin{equation}
\alpha_{12} = \frac{q_{\rm 1p} q_{\rm 2p} D_{12}}{\hbar c} =
\frac{q_{\rm 1p} q_{\rm 2p}}{\hbar c}\left(\frac{q_{\rm 1a}}{q_1}
+ \frac{m_2}{M} C_{21}\right) \, .
\end{equation}
Therefore, a comparison of the energy levels in atoms having
different nuclear mass yields a test of the equality of active and
passive charges. Since the accuracy is influenced by the accuracy
of the theoretical prediction of the transition frequencies,
comparison of simple atoms gives the most accurate results. Let us
thus compare hydrogen H and ionized helium He$^+$: For H, we have
$q_1 = q_p$, $q_2 = q_e$, and $M\approx m_p$; for He$^+$, $q_1 = 2
q_p$ and $q_2 = q_e$, and $M\approx 4m_p$. For the ratio of the
transition frequencies, we thus find
\begin{equation}
\frac{\nu_{12}({\rm He}^+)}{4\nu_{12}({\rm
H})}=\frac{\alpha_{12}^2({\rm He}^+)}{4\alpha_{12}^2({\rm
H})}\approx 1- \frac 32 \frac{m_e}{m_p} C_{21} \label{AtomC12}
\end{equation}
where we have neglected terms of the order $(m_e/m_p)^2$. The
1S$_{1/2}$-2S$_{1/2}$ transition in hydrogen has been measured to
a precision of $1.9\times 10^{-14}$ \cite{Udemetal97}; however,
the theoretical prediction of the Lamb shift has an error bar of
$6.9\times 10^{-13}$, in part due to the uncertainty in the charge
radius of the proton. See Appendix A of Ref. \cite{MohrTaylor05}.
Since He$^+$ ions can be laser cooled \cite{Roth}, an even higher
precision is expected for them \cite{Haenschxx}. Also the
theoretical uncertainty for He$^+$ can be lower since the
properties of the He nucleus are better known. If no discrepancy
like Eq.~(\ref{AtomC12}) between H and He$^+$ at $~7\times
10^{-13}$ would be found, we could deduce a limit of $|C_{21}|
\leq 8.3 \times 10^{-10}$.

The accuracy of this limit is less than the one of the frequencies
by the nucleus to electron mass ratio of hydrogen $m_p/m_e$. Thus,
it is interesting to consider positronium, where the mass ratio is
unity. Positronium's $1\,^3S_1-2\,^3S_1$ frequency is known to
$2.6\times 10^{-9}$ \cite{FeeChu}. The deviation from the
theoretical prediction is $\sim 1.3\sigma$ (\cite{Karshenboim},
Fig. 3), but we consider this insignificant. Comparison to
hydrogen thus yields a limit on
$\alpha_{ee^+}^2/\alpha_{12}^2({\rm H}) \approx 1+ C_{e\rm e^+}$,
where we neglect a term which is suppressed by $m_e/m_p$. Thus, we
obtain $|C_{\rm e\rm e^+}|\leq 2.6\times 10^{-9}$ (The Lamb shift
in positronium can be predicted to sufficient accuracy
\cite{MohrTaylor05}). These limits are not as precise as the ones
derived from bulk matter experiments, which gain sensitivity from
the macroscopic number of particles. However, they are
particularly clean: The physics of light atoms is known in detail
to the same precision as the limit derived.

The explicit form of the experimental signature Eq.
(\ref{AtomC12}) also makes clear that the measurement is not
insensitive to changes in the passive charge even if it is
accompanied by changes in the inertial masses. These masses solely
enter the factor $m_e/m_p$ that sets the sensitivity of the
experiment to $C_{21}$. However, $C_{21}\neq 0$ will always be
detected, regardless of small variations in this factor. The
fundamental reason for this is that the experiment is based on
comparing two atoms that have different nuclei and thus different
charge to mass ratios.

\section{Magnetic moments}

Since moving charges create magnetic fields and magnetic fields
act on moving charges, one may extend the above analysis to the
question of the equality of active and passive magnetic moments.
In analogy to the above considerations, we first calculate the
force between two magnetic moments
\begin{eqnarray}
m_1 \ddot{\mbox{\boldmath$x$}}_1 & = &
{\mbox{\boldmath$\nabla$}}_1 \left({\mbox{\boldmath$\mu$}}_{1 \rm
p} \cdot {\mbox{\boldmath$B$}}_2({\mbox{\boldmath$x$}}_1)\right)
\,,\nonumber \\ m_2 \ddot{\mbox{\boldmath$x$}}_2 & = &
{\mbox{\boldmath$\nabla$}}_2 \left({\mbox{\boldmath$\mu$}}_{2 \rm
p} \cdot {\mbox{\boldmath$B$}}_1({\mbox{\boldmath$x$}}_2)\right)
\,,
\end{eqnarray}
where
\begin{equation}
{\mbox{\boldmath$B$}}_j({\mbox{\boldmath$x$}}_k)=\frac{3 (({\mbox{\boldmath$x$}}_k - {\mbox{\boldmath$x$}}_j) \cdot {\mbox{\boldmath$\mu$}}_{j \rm a}) ({\mbox{\boldmath$x$}}_k - {\mbox{\boldmath$x$}}_j) - {\mbox{\boldmath$\mu$}}_{j \rm a}|{\mbox{\boldmath$x$}}_k - {\mbox{\boldmath$x$}}_j|^2}{|{\mbox{\boldmath$x$}}_k - {\mbox{\boldmath$x$}}_j|^5}\,.
\end{equation}
In a classical picture, the magnetic moments can be considered as
being created by a current loop and the direction of the magnetic
moment is given by the orientation of the loop. Therefore, a
difference between active and passive magnetic moments is a
difference between their magnitudes only, which are related to the
charges making up the current. Thus, we assume
${\mbox{\boldmath$\mu$}}_{1,2 \rm a,p} = \mu_{1,2 \rm a,p}
\widehat{\mbox{\boldmath$\mu$}}_{1,2}$, where
$\widehat{\mbox{\boldmath$\mu$}}_{1,2}$ are unit vectors
indicating the direction of the magnetic moments.  If we introduce
\begin{eqnarray}
\widetilde C_{21} &=& \frac{\mu_{2 \rm a}}{\mu_{2 \rm p}} - \frac{\mu_{1 \rm a}}{\mu_{1 \rm p}} \, ,\nonumber \\
\widetilde D_{21} &=& \frac{m_1}{M} \frac{\mu_{1 \rm a}}{\mu_{1
\rm p}} + \frac{m_2}{M} \frac{\mu_{2 \rm a}}{\mu_{2 \rm p}} =
\frac{\mu_{1 \rm a}}{\mu_{1 \rm p}} + \frac{m_2}{M} \widetilde
C_{21}
\end{eqnarray}
then we obtain for the center--of--mass and relative coordinates
\begin{align}
\ddot{\mbox{\boldmath$X$}} & = - \frac{\mu_{2 \rm p} \mu_{1 \rm
p}}{M} \widetilde C_{21} \mbox{\boldmath$\nabla$} \frac{3
(\mbox{\boldmath$x$} \cdot \widehat{\mbox{\boldmath$\mu$}}_2)
(\widehat{\mbox{\boldmath$\mu$}}_1 \cdot \mbox{\boldmath$x$}) -
\widehat{\mbox{\boldmath$\mu$}}_1 \cdot
\widehat{\mbox{\boldmath$\mu$}}_2
|\mbox{\boldmath$x$}|^2}{|\mbox{\boldmath$x$}|^5}\,,\nonumber
 \\
\ddot{\mbox{\boldmath$x$}} & = \frac{\mu_{1 \rm p} \mu_{2 \rm p}}{m_{\rm red}}
\widetilde D_{21} \mbox{\boldmath$\nabla$} \frac{3 (\mbox{\boldmath$x$} \cdot
\widehat{\mbox{\boldmath$\mu$}}_2) (\widehat{\mbox{\boldmath$\mu$}}_1 \cdot
\mbox{\boldmath$x$}) - \widehat{\mbox{\boldmath$\mu$}}_1 \cdot \widehat{\mbox{\boldmath$\mu$}}_2
 |\mbox{\boldmath$x$}|^2}{|\mbox{\boldmath$x$}|^5} \, .
\end{align}
For different ratios of active and passive magnetic moments, the
center of mass will show self--acceleration. The equations of the
torque describe the orientation of the magnetic moments only: $
\dot{\mbox{\boldmath$\mu$}}_{1 \rm p} = -
{\mbox{\boldmath$B$}}_2({\mbox{\boldmath$x$}}_1) \times
{\mbox{\boldmath$\mu$}}_{1 \rm p}$, and $
\dot{\mbox{\boldmath$\mu$}}_{2 \rm p} = -
{\mbox{\boldmath$B$}}_1({\mbox{\boldmath$x$}}_2) \times
{\mbox{\boldmath$\mu$}}_{2 \rm p}$. This gives rise to an
additional spin--orbit coupling. We will not consider this.

\subsection{Atomic spectroscopy}

Again, the relative motion can contribute to the energy of a
hydrogen atom: the Hamiltonian for the hyperfine interaction reads
\begin{eqnarray}
H_{\rm hf} & = & - \frac{\mu_{1 \rm p} \mu_{2 \rm p}}{m_{\rm red}}
\widetilde D_{21} \left(\frac{8 \pi}{3} \delta(x)
\widehat{\mbox{\boldmath$\mu$}}_1 \cdot
\widehat{\mbox{\boldmath$\mu$}}_2 \nonumber \right. \\ & & \left.
+ \frac{3 (\mbox{\boldmath$x$} \cdot
\widehat{\mbox{\boldmath$\mu$}}_2)
(\widehat{\mbox{\boldmath$\mu$}}_1 \cdot \mbox{\boldmath$x$}) -
\widehat{\mbox{\boldmath$\mu$}}_1 \cdot
\widehat{\mbox{\boldmath$\mu$}}_2
|\mbox{\boldmath$x$}|^2}{|\mbox{\boldmath$x$}|^5}\right)\label{DeltaHFS}
\end{eqnarray}
where the $\delta$--function describes the contribution from the
local interaction of the electron with the nucleus. The
$\widehat{\mbox{\boldmath$\mu$}}_2$ are now total angular momentum
operators.

To obtain experimental limits, we compare the hyperfine splitting
of atoms having different nuclei. The hyperfine splitting of
muonium has been measured to an accuracy of $1.1 \times 10^{-8}$
\cite{Liu}. As summarized in \cite{KarshenboimIvanov02}, it is
compatible with the theoretical prediction, which has an
uncertainty of $1.2\times 10^{-7}$. For positronium, two precision
measurements of the 1S hyperfine splitting have been reported by
Mills {\em et al.} \cite{Mills}, and Ritter {\em at al.}
\cite{Ritter}. They agree within the experimental error. However,
they deviate by -6.3(2.9)\,MHz and -4.7(1.7)\,MHz, respectively,
from the theoretical prediction of 203.3917(6)\,GHz
\cite{Karshenboim}.

In our framework, this difference may be modelled by a difference
of passive and active magnetic moments of $\widetilde C_{ee^+}=
-1.7(7)\times 10^{-5}$ for the Mills {\em et al.} measurement and
$\widetilde C_{\rm ee^+}= -1.3(4)\times 10^{-5}$ for Ritter {\em
et al.}. It is worth noting that such a discrepancy between theory
and experiment exist for this 1S hyperfine splitting only. All
other spectroscopical quantities discussed in \cite{Karshenboim},
which are approximately independent of a difference in active and
passive magnetic moments, agree to their theoretical prediction
(e.g., the 1S-2S interval already used above to find a limit on
$C_{\rm ee+}$ and the fine structure of four different
transitions).

On the other hand, the discrepancy between theory and experiment
could be due to systematic influences in the experiments or an
incomplete theoretical understanding of positronium (the latter
point of view is expressed in \cite{Karshenboim}). Even two
different experiments can be influenced by the same systematic
effects if the measurement principle is similar or in part
similar. For example, the line shape of positronium is still being
investigated \cite{Hughes}.

This would make it interesting to find alternative measurements,
for example from the hyperfine structure of non-leptonic atoms. An
overview for hydrogen, deuterium, tritium and the $^3$He ion can
be found in \cite{KarshenboimIvanov02}. Precise experiments (with
error bars in the $10^{-12}$ range) do exist, but unfortunately
there are rather large discrepancies to the theory. These are
attributed to the uncertainty of the nuclear contributions
\cite{KarshenboimIvanov02}. Moreover, the high nucleus to electron
mass ratio suppresses the influence of $\widetilde C_{12}$ on the
hyperfine splitting of non-leptonic atoms. For example, hydrogen
and tritium show the lowest discrepancy between theory and
experiment of -33ppm and -38ppm, respectively. We take the
difference of 5\,ppm as a signal for active and passive magnetic
moments, and the geometric sum of 50\,ppm of the discrepancies as
an estimate of the error. This gives $ |\widetilde C_{21}({\rm
H})-\frac 13\widetilde C_{21}({\rm T})|\leq \frac12
\frac{m_p}{m_e}\times (5\pm50){\rm ppm}=0.005\pm 0.045$. This is
not suitable for ruling out the significant value from positronium
and muonium spectroscopy.

To sum up, at the present state of theory and experiment we have
to regard the possibility of a difference of theory and experiment
as a hypothesis that is probably wrong, though it is supported by
two independent experiments. However, it would be interesting to
check it against other systems. Unfortunately, the hyperfine
structure of non-leptonic atoms depicts even larger theoretical
uncertainties.

\section{Summary and outlook}

In summary, we have introduced the concept of active and passive
electric charges and magnetic moments, which in standard
electrodynamics are assumed to be equal. The best limits (of the
order of $10^{-21}$ for $C_{pe}$ come from experiments testing the
neutrality of macroscopic matter, which gain sensitivity from a
large number of particles. Spectroscopy of hydrogen and
positronium provides $|C_{ee^+}|\leq 2.6\times 10^{-9}$. These
limits can also be interpreted as experimental verifications of
Newton's third law for electric forces at the $10^{-21}$ level.
For magnetic moments, comparison of the hyperfine structure of
positronium and muonium suggests a difference of active and
passive of $|\tilde C_{ee^+}| \leq -1.4\times 10^{-5}$ that is
significant at the $3\sigma$ level. However, at the present state
of theory and experiment, this is more likely an artifact, even if
two independent experiments agree. The best limit from
non-leptonic atoms (hydrogen and tritium) is at the 5\% level of
accuracy.

The relativistic quantum theory of electrodynamics is quantum
electrodynamics (QED), and one may ask how the question of active
and passive charges can be formulated in this context. The most
straightforward way to do so starts at the level of the field
equations: We use the passive charge in the Dirac equation for an
electron with minimal coupling to the electromagnetic field. The
active charge enters the source term in the inhomogenous Maxwell
equation. The non-relativistic Pauli equation and the classical
limit can then be found in the usual way, as described in
textbooks. As a result, this relativistic quantum description
contains our above classical results. Further experimental
signatures on a difference of active and passive charges might be
sought within such a model. This might be possible, for example,
by comparing the value of the fine-structure constant obtained
from the measurement of the electron's anomalous magnetic moment
$g-2=\alpha/(2\pi)+\ldots$ (to $7\times 10^{-10}$ accuracy
\cite{Gabrielse2006}), to other measurements of $\alpha$. However,
even without a QED version of our question, we were able to answer
it in terms of experimental limits, some of them very stringent.

The $C$ and $\tilde C$ coefficients for most particle combinations
are $\sim 12$ orders of magnitude less stringent than $C_{pe}$ and
$C_{ne}$. Thus, it might be interesting to seek further
experimental limits. For example, certain selection rules in
spectroscopy, that are normally imposed by symmetry arguments,
might be broken. New versions of the macroscopic matter
experiments could simultaneously measure the active and passive
charge in order to suppress some of the systematic effects.

We like to thank S. Chu, H. Dittus, F.W. Hehl, K. Jungmann, and
C.S. Unnikrishnan for fruitful discussions. Financial support from
the German Space Agency DLR and CONACyT grant 48404--F is
acknowledged. H.M. thanks the Alexander von Humboldt Foundation
for their support.

\end{document}